# Modelling Carbon Coated Silicon Anodes for Lithium-Ion Batteries and the Influence of Internal Contact Area on Rate Performance

**Dharshannan Sugunan** [1], Yang Jiang [1,2], Jia Guo [1], Huizhi Wang [1], Monica Marinescu [1], Gregory J. Offer [1,2,*,z]

[1] Imperial College London, London, UK, [2] Breathe Battery Technologies, London, UK

## Abstract

Silicon is a promising anode material for next-generation lithium-ion batteries due to its exceptionally high specific capacity (~3600 mAh g⁻¹), significantly exceeding that of conventional graphite. However, its practical application is hindered by substantial volume expansion (300–400%) during lithiation, leading to mechanical degradation and capacity fade. Graphite-coated silicon core–shell structure have been developed to mitigate these issues by combining silicon's capacity with graphite's structural stability. Despite this, experimental studies have shown that the usable capacity of such composite electrodes can remain low, especially under high-rate cycling, often below 40% at 1 C. In this work, we develop a physics-based electrochemical model in PyBaMM to investigate the charge–discharge behaviour, rate limitations, and degradation mechanisms of silicon–graphite core–shell anodes. The model incorporates lithium transport, interfacial kinetics, evolving contact area due to silicon expansion, and a simplified cracking framework to capture loss of active material. Results are validated against key experimental trends and used to explore the effects of particle size, shell thickness, and charge protocol, offering insights into the design of more durable and efficient Si-based composite anodes.

## 1. Introduction

Silicon offers a theoretical capacity of ~3600 mAh g⁻¹, nearly ten times higher than that of graphite (~372 mAh g⁻¹), making it a promising anode material for lithium-ion batteries [1]. However, its large volume expansion during lithiation (300–400%) leads to mechanical degradation and capacity fade [2]. Core–shell structures combining silicon with graphite aim to address this, but studies report low utilization, often below 40% at 1 C, due to kinetic limitations and structural inhomogeneity [3]. Electrochemical modelling is therefore essential to better understand and improve the performance of these composite electrodes.

In traditional silicon-graphite composite electrodes, the lithiation and delithiation processes of silicon and graphite occur independently. This independence fails to effectively address the degradation issues of silicon negative electrodes. For instance, during lithiation, silicon undergoes significant volume expansion, leading to particle fracture and repeated breaking and reformation of the solid electrolyte interphase (SEI), thereby accelerating negative electrode degradation and the occurrence of side reactions [4,5]. In contrast, the silicon in core-shell structures is encapsulated within a carbon shell, effectively mitigating side reactions caused by exposed silicon particles. However, these structures extends the lithium-ion transport pathway, and complicates the reaction kinetics. Depending on how the core-shell structures are made, they may or may not have internal void spaces to accommodate silicon expansion. If they do not, then mechanical stress will build up internally, until particles fracture under strain. At this point silicon can become exposed to electrolyte and accelerated side reactions will occur.

During negative electrode lithiation, lithium ions first intercalate into the carbon before reaching the silicon core. Consequently, the electrochemical potential of the silicon core remains higher than that of the carbon shell until the negative electrode is fully lithiated. For core-shell structures without internal void spaces, the mechanical stress caused by the volume expansion of silicon against the carbon can become the main source of the potential difference between the carbon and the silicon. This phenomenon has been demonstrated and modeled by Roper et al. [6]. However, when sufficient space is left between the carbon shell and the silicon core, the expansion of the silicon can be accommodated by the void, so the mechanical stress is no longer a significant source of potential difference between silicon and carbon.

Although core-shell structured silicon-carbon negative electrodes with voids may avoid the instability caused by mechanical stress, they do not always seem to significantly improve capacity. This reduction in capacity becomes particularly evident at higher charge and discharge rates, where the utilization of silicon is significantly limited. Among the reported studies, the largest rate-dependent capacity loss is observed in the work by Wang et al. [7], which investigated M-Si@void@C negative electrodes. These electrodes consist of carbon-coated silicon particles containing internal voids, with the "M" indicating preparation using a medium concentration of poly(diallyldimethylammonium chloride) (PDDA)-modified silicon. As the delithiation current increased from 0.1 A $g^{-1}$ to 10.0 A $g^{-1}$, the capacity dropped significantly from 2200 mAh $g^{-1}$ to 400 mAh $g^{-1}$.

Similarly, Pan et al. [8] reported a capacity decrease from 1560 mAh $g^{-1}$ to approximately 700 mAh $g^{-1}$ as the delithiation current increased from 0.1 A $g^{-1}$ to 1.0 A $g^{-1}$ using Si@void@C composite electrodes. In another study, [9] observed that a core–shell silicon–carbon negative electrode, composed of 84.1 percent silicon and 15.9 percent carbon by mass, exhibited a drop in utilization rate from 73.5 percent at 0.032 C to 50.4 percent at 0.32 C.

In comparison, [10] reported a more moderate capacity loss, with values decreasing from 1580 mAh $g^{-1}$ to 1100 mAh $g^{-1}$ as the delithiation current increased from 0.2 C to 10 C. Notably, [10] employed smaller nanoscale silicon particles (10-50 nanometres), whereas [7] used larger particles (50-150 nanometres), highlighting the influence of particle size on capacity retention. Smaller particles offer a larger surface area and improved contact with the carbon matrix, which would enhance interfacial reaction rates and mitigate diffusion limitations under high current conditions.

These results collectively underscore the importance of current density, silicon particle size, and interfacial contact in determining the rate capability and utilization of silicon-based composite anodes, a relationship that will be further examined through the modelling results in the following sections. Although concentration gradients can contribute to polarization, the observed voltage rise in the core–shell structure appears more abrupt than that of electrodes composed of pure carbon or pure silicon. Additionally, the decrease in utilization with increasing C-rate suggests the presence of a kinetic limitation. We hypothesize that a chemical potential barrier exists at the carbon–silicon interface. Since lithium transfer between the shell and core is a chemical process rather than an electrochemical one, this barrier is likely driven by a difference in chemical potential rather than by an electrochemical overpotential.

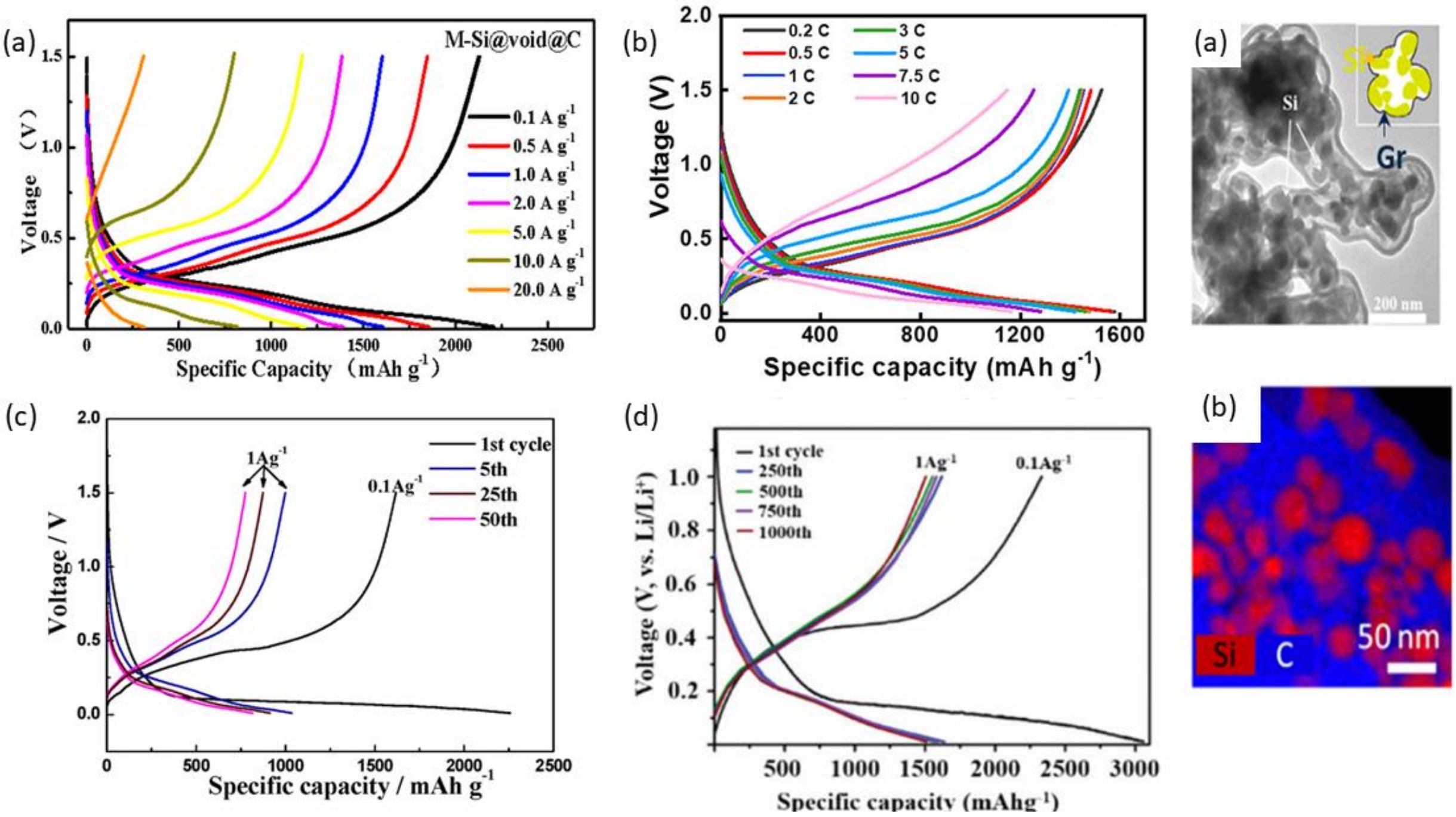

Fig. 1 (a) Charge and discharge curves of the M-Si@void@C negative electrode under different current densities[7]; (b) corresponding voltage profiles between 0.1 and 10 C[10]; (c) discharge–charge profiles of the Si@Void@C negative electrode at 0.1 $Ag^{-1}$ and 1 $Ag^{-1}$ current [8]; (d) voltage profiles plotted for the battery discharge/charge at 1 $Ag^{-1}$ and 0.1 $Ag^{-1}$ [9]. Reproduced with permission from the journals as cited in [7–10].

In this paper, we employ dilute solution theory to calculate the chemical potential difference between the carbon shell and the silicon core, combined with an electrochemical model to simulate the charge and discharge (lithiation–delithiation) behaviour of the core–shell structured negative electrode. The model also accounts for changes in the contact area between the core and shell due to the volume expansion and contraction of silicon during cycling, which affects the lithium-ion diffusion flux across the interface. The model is qualitatively validated against experimental observations; quantitative validation is deferred due to the need for detailed experimental measurements and precise parameter fitting. Finally, a degradation model is developed to describe the degradation mechanisms of the cell.

Studies from [8] and [9] also account for cycling-induced capacity loss due to degradation, where we theorize that the void structure cannot entirely accommodate the substantial volume expansion of silicon resulting in cracking as the primary degradation mechanism. While the first few sections focus on core–shell dynamics and interface kinetics, a dedicated degradation model is presented in a later section to capture these effects more explicitly.

## 2. Model Development

This work models the battery charging and discharging behaviours of a core-shell structured Si/C composite negative electrode. The core-shell model is developed based on the single particle model (SPM) framework. Model assumptions are as follows.

- Silicon is isolated from the electrochemistry of the system.
- The modelled carbon shell is graphitic.
- Butler-Volmer kinetics applied to graphite shell surface.
- $Li$ transfer between graphite and silicon is governed by an interfacial chemical reaction.

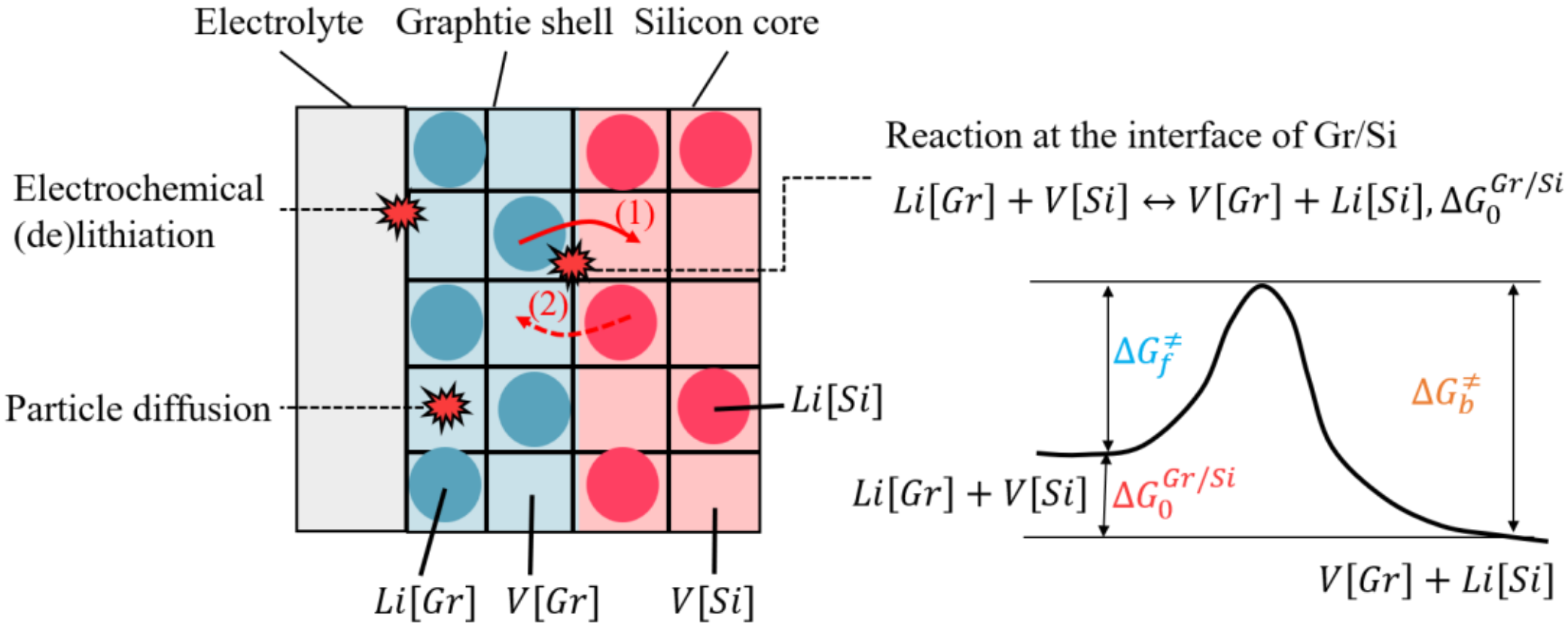

Fig. 2 Schematic of the core-shell structure of a graphite coated silicon particle.

As illustrated in the schematic in Figure 2, the model incorporates three main processes: (1) electrochemical (de)lithiation between the electrolyte and the graphite shell, (2) lithium diffusion within both the graphite and silicon phases, and (3) a chemical reaction at the interface between the graphite shell and the silicon core. The overall mechanism can be considered a 4-step electrochemical-diffusion-chemical-diffusion process (EC-D-C-D), compared to a normal model for a single particle being simply a 2-step electrochemical-diffusion (EC-D) process.

## 2.1 Model Equations

The surface electrochemical reaction can be defined by:

$$Li^{+}[electrolyte] + e^{-}[solid] + V[Gr] \Leftrightarrow Li[Gr],\ \Delta G_0^{Gr} \tag{1}$$

Where "[]" refers to the corresponding phases; $V[Gr]$ means the vacant sites for Gr; $Li[Gr]$ refers to the lithiated Gr phase; $\Delta G_0^{Gr}$ (J mol$^{-1}$) denotes the standard Gibbs free energy change associated with Li+ intercalation into the Gr phase, governing the equilibrium potential of the reaction. The equilibrium potential of this reaction is determined by the lithiation status in Gr:

$$\begin{aligned} E_{eq}^{Gr} &= 0.7222 + 0.1387X_{Li[Gr]} + 0.029X_{Li[Gr]}^{0.5} \\ &\quad - \frac{0.0172}{X_{Li[Gr]}} + \frac{0.0019}{X_{Li[Gr]}^{1.5}} \\ &\quad + 0.2808e^{0.9-15X_{Li[Gr]}} - 0.7984e^{0.4465X_{Li[Gr]}-0.4108} \end{aligned} \tag{2}$$

The relationship between $E_{eq}^{Gr}$ (V) and $\Delta G_0^{Gr}$ is (similar for silicon):

$$\Delta G_0^{Gr} = -nFE_{eq}^{C} \tag{3}$$

Where n = 1, and $F$ (96485 C mol$^{-1}$) is the Faraday Constant. The lithiation status $X_{Li[Gr]}$ is determined by:

$$X_{Li[Gr]} = \frac{C_{Gr}}{C_{Gr}^{max}} \tag{4}$$

Where $C$ and $C_{Gr}^{max}$ (mol m$^{-3}$) refer to the lithium concentration and the maximum concentration of Gr.

Following the similar way in equation (2) and (3), we have equilibrium potential for Si ($E_{eq}^{Si}$, V) where lithiation and delithiation potentials are separately fitted and the corresponding standard Gibbs free energy change $\Delta G_0^{Si}$ (J mol$^{-1}$) determined:

$$\begin{aligned} E_{eq,lith}^{Si} &= -96.63X_{Li[Si]}^{7} + 372.6X_{Li[Si]}^{6} - 587.6X_{Li[Si]}^{5} + 489.9X_{Li[Si]}^{4} \\ &\quad - 232.8X_{Li[Si]}^{3} + 62.99X_{Li[Si]}^{2} - 9.286X_{Li[Si]} + 0.86333 \\ E_{eq,delith}^{Si} &= -51.02X_{Li[Si]}^{7} + 161.3X_{Li[Si]}^{6} - 205.7X_{Li[Si]}^{5} + 140.2X_{Li[Si]}^{4} \\ &\quad - 58.76X_{Li[Si]}^{3} + 16.87X_{Li[Si]}^{2} - 3.792X_{Li[Si]} + 0.9937 \end{aligned} \tag{5}$$

$$X_{Li[Si]} = \frac{C_{Si}}{C_{Si}^{max}} \tag{6}$$

The Butler-Volmer equation is used to describe the electrochemical reaction at the interface between electrolyte and graphite surface:

$$\eta^{Gr} = \frac{2RT}{F} asinh\left(\frac{i}{2i_0^{Gr}}\right) \tag{7}$$

where $\eta^{Gr}$ (V) is the overpotential, $T$ (K) is the cell temperature (assumed to be 298.15 K unless otherwise specified); $i$ is the applied current density (A m$^{-2}$); and $i_0^{Gr}$ (A m$^{-2}$) denotes the exchange current density of the graphite phase:

$$i_0^{Gr} = i_{00}^{Gr}\left(1 - X_{Li[Gr]}\right)^{\alpha}\left(X_{Li[Gr]}\right)^{1-\alpha} \tag{8}$$

where $i_{00}^{Gr}$ (A m$^{-2}$) is the reference exchange current density for Gr:

$$i_{00}^{Gr} = k^{Gr} F C_{Gr}^{max}\left(1 - X_{Li[Gr]}\right)^{\alpha}\left(X_{Li[Gr]}\right)^{1-\alpha} \tag{9}$$

Where $k^{Gr}$ (m$^{2.5}$ mol$^{-0.5}$ s$^{-1}$) is the intercalation rate constant of Gr, and $\alpha$ is the charge transfer coefficient (assumed to be 0.5 in this work).

The surface electrode potential of Gr is given by:

$$\phi^{Gr} = E_{eq}^{Gr} - \eta^{Gr} \tag{10}$$

Hence, the output voltage $V$ (V) can be determined by the surface potential $\phi^{Gr}$:

$$V = \phi^{Gr} \tag{11}$$

The particle diffusion inside both Gr and Si phases is governed by:

$$\frac{\partial C}{\partial t} = \frac{1}{r^2}\frac{\partial}{\partial r}\left(D_m r^2 \frac{\partial C}{\partial r}\right), m = Gr, Si \tag{12}$$

And two boundary conditions apply to each phase as:

$$\left.\frac{\partial C}{\partial r}\right|_{r=R_{Gr}} = \frac{-i}{D_{Gr}F} \tag{13}$$

$$A_{Gr}D_{Gr}\left.\frac{\partial C}{\partial r}\right|_{r=R_{Gr/Si}} = A_{Cont}N_{Gr/Si} \tag{14}$$

$$A_{Cont}D_{Si}\left.\frac{\partial C}{\partial r}\right|_{r=R_{Gr/Si}} = -A_{Cont}N_{Gr/Si} \tag{15}$$

$$\left.\frac{\partial C}{\partial r}\right|_{r=R_{Gr/Si}-d} = 0 \tag{16}$$

Where $D_m$ (m=Gr, Si) denotes the diffusion coefficient in each phase. Considering the microstructure of the Gr-coated Si architecture, as described in Section 2.2, the inner surface area of the graphite shell $A_{Gr}$ (m$^2$) is not necessarily equal to the contact area with the silicon core, $A_{cont}$. The effective diffusion length $d$ (m) within the silicon core is also defined in Section 2.2.

The lithium flux at the graphite–silicon interface ($N_{Gr/Si}$) is governed by the interfacial chemical reaction, which is expressed as the difference between the forward and backward reaction fluxes (mol m$^{-2}$ s$^{-1}$), based on transition state theory:

$$N_{Gr/Si} = N_{Gr2Si} - N_{Si2Gr} \tag{17}$$

The forward and backward reaction fluxes $N_{Gr2Si}$ and $N_{Si2Gr}$ are proportional to the concentrations of the respective reactants:

$$N_{Gr2Si} = k_f\left(1 - X_{Li[Si]}\right)X_{Li[Gr]} \tag{18}$$

$$N_{Si2Gr} = k_b X_{Li[Si]}\left(1 - X_{Li[Gr]}\right) \tag{19}$$

To the best of our knowledge, there are no existing measurements of the forward and backward reaction rate constants $k_f$ and $k_b$ for this specific reaction. However, according to equilibrium theory, they are expected to follow the relationship below:

$$k_{eq} = \frac{k_f}{k_b} = e^{-\Delta G_0^{Gr/Si}} \tag{20}$$

Although $\Delta G_0^{Gr/Si}$ has not been experimentally determined, the Gibbs free energy difference between the two steady states can be inferred by applying the principle of superposition, as proposed in our work:

$$Li[Gr] + V[Si] \leftrightarrow V[Gr] + Li[Si], \Delta G_0^{Gr/Si} \tag{21}$$

The overall chemical reaction above can be decomposed into the sum of two electrochemical reactions: Eqn. (1) and the following equation:

$$Li^{+}[electrolyte] + e^{-}[solid] + V[Si] \Leftrightarrow Li[Si],\ \Delta G_0^{Si} \tag{22}$$

Hence, the $\Delta G_0^{Gr/Si}$ can be determined by:

$$\Delta G_0^{Gr/Si} = \Delta G_0^{Si} - \Delta G_0^{Gr} \tag{23}$$

By doing so, it can be shown that $k_{eq}$ is determined by the equilibrium potentials $E_{eq}^{Gr}$ and $E_{eq}^{Si}$, and the relationship between $k_f$ and $k_b$ is accordingly established. Within this modelling framework, only one of the two parameters—either $k_f$ or $k_b$ remains as an independent variable. The parameters values are shown in Table S1 and variables solved for in Table S2.

## 2.2 Core-Shell Model Structure

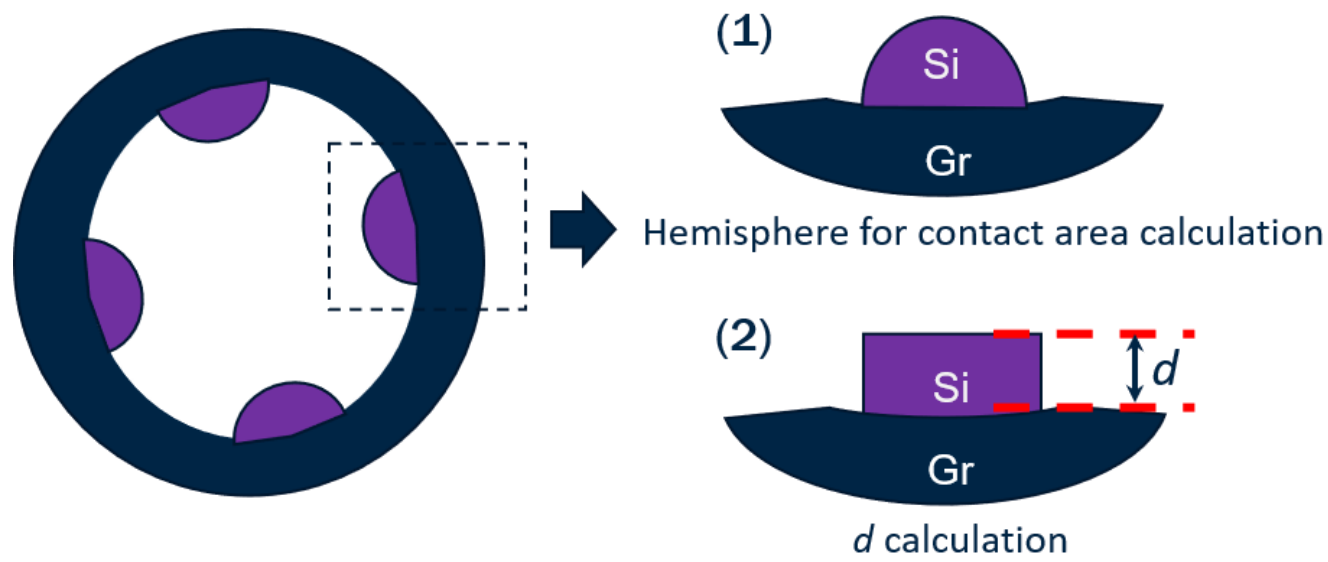

Fig. 3 Contact between graphite shell and Silicon core.

The silicon core is assumed to be attached to the inner surface of the graphite shell in the form of a hemisphere, with the contact area approximated by the base area of that hemisphere. Given that silicon can undergo volumetric expansion of up to 300% during lithiation, while graphite expands by only around 10%, the volume of the graphite shell is considered effectively constant. As a result, the dynamic volume change of silicon during charging and discharging leads to a corresponding evolution in the interfacial contact area, which directly impacts the lithium transport and reaction kinetics at the interface.

$$V_0 = \frac{2}{3}\pi R_0^3 \tag{24}$$

When the volume of the hemisphere increases from $V_0$ to $V$, the surface area of the hemisphere will increase from $\pi R_0^2$ to $\pi R^2$. Since the change in the volume of silicon is a measurable parameter, the contact area will evolve from an initial $A_{Cont,0}$ to $A_{Cont}$, given by:

$$A_{Cont,0} = (\frac{9\pi}{4})^{1/3}(V_0)^{2/3} \tag{25}$$

$$A_{Cont} = (\frac{9\pi}{4})^{1/3}(V)^{2/3} \tag{26}$$

$$A_{Cont}/A_{Cont,0} = (\frac{V}{V_0})^{2/3} \tag{27}$$

Here, the ratio $V/V_0$ is derived from experimental test results in the literature and varies with the silicon potential. The thickness of silicon can be calculated based on an equivalent cylinder:

$$d_{Si} = \frac{V}{A_{Cont}} = (\frac{4}{9\pi})^{1/3} \cdot V^{1/3} \tag{28}$$

$$d_{Si} \leq 1/3R \tag{29}$$

## 2.3 Silicon Moving Boundary Formulation

The base model is further modified to incorporate a moving boundary formulation that captures the volume change of the silicon core enclosed by the graphite shell. To facilitate this, the geometry of the model is simplified, as illustrated in Figure 4.

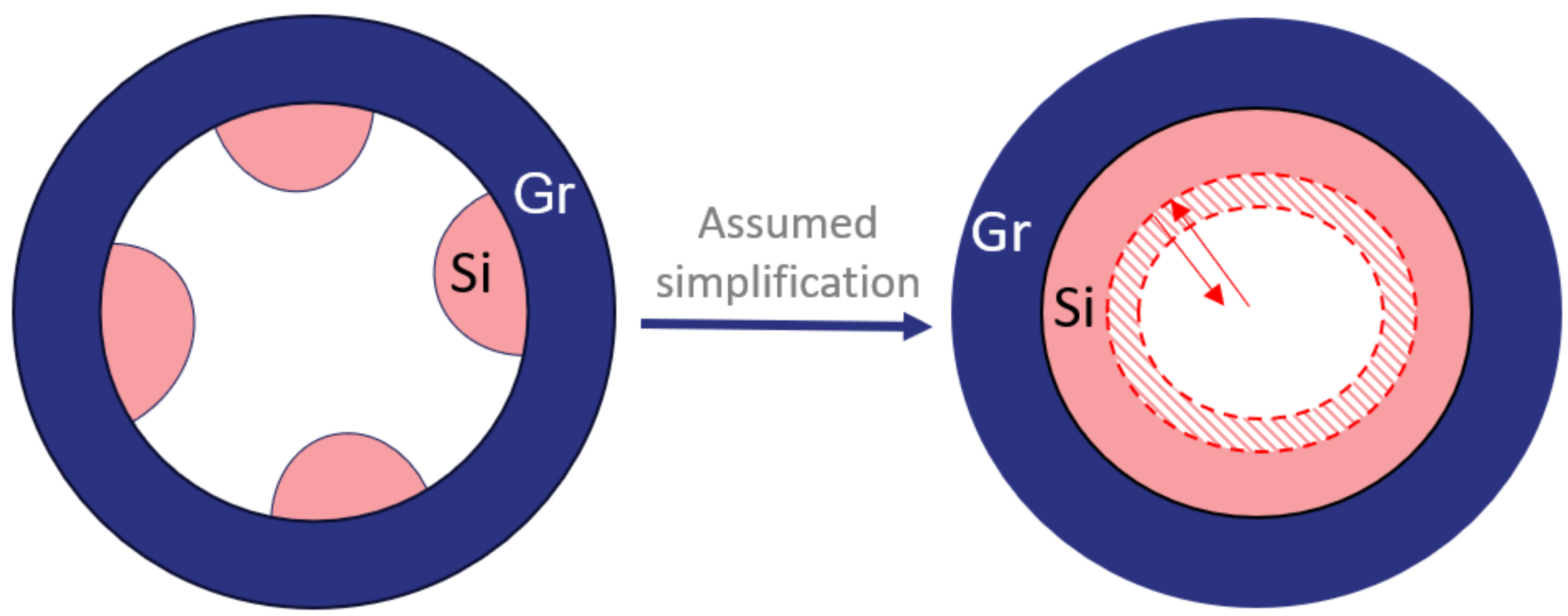

Fig. 4 Model structure simplification to allow for moving boundary implementation.

In this simplification, the initially proposed structure (left) is approximated by assuming that the silicon core consists of infinitesimally small particles uniformly distributed along the inner surface of the graphite shell. Under this assumption, the silicon core is represented as a thick ring (right) rather than a hemispherical particle. However, the volume change and surface area evolution of the silicon are still computed using the expressions derived for a hemispherical geometry, as given by equations (24–29). The inner surface of the silicon (highlighted in red) is modelled as a free-moving boundary to account for the volumetric expansion and contraction of silicon during lithiation and delithiation.

The moving boundary is incorporated through a coordinate transformation that normalizes both the graphite and silicon domains. This transformation is applied to the diffusion equations in each domain to account for the evolving geometry. The transformed equations for the silicon domain are presented below. The graphite domain follows the same formulation, but in a simplified form due to the assumption of no volume change.

Silicon domain normalization:

$$r_i = r_o - d_{si} \tag{30}$$

$$\xi = \frac{r - r_i(\phi)}{r_o - r_i(\phi)}, \quad 0 \le \xi \le 1 \tag{31}$$

$$\frac{\partial}{\partial r} = \frac{1}{r_o - r_i(\phi)} \frac{\partial}{\partial \xi} \tag{32}$$

$$\frac{\partial}{\partial t} = \frac{\partial}{\partial t} + \frac{\partial \phi}{\partial t}\frac{\partial}{\partial \phi} - \frac{\partial r_i}{\partial \phi}\frac{\partial \phi}{\partial t}\left(\frac{\xi}{r_o - r_i(\phi)}\right)\frac{\partial}{\partial \xi} \tag{33}$$

The parameter $d_{si}$ corresponds to the silicon thickness as defined in Equation (29). The variables $r_i$ & $r_o$ represent the inner (moving) and outer (fixed) radii of the thick silicon core ring, respectively. $\phi$ here is the **surface stoichiometry** of the silicon particle. The volume change of silicon, $\frac{V}{V_0}$, is fitted using experimental data reported by Uxa et al. [11].The complete set of equations governing the moving boundary evolution is presented below.

Full definition of silicon domain with moving boundary incorporation:

$$\frac{V}{V_0} = f(\phi) \tag{34}$$

$$\frac{\partial C}{\partial t} + \frac{\partial \phi}{\partial t}\left(\frac{\partial C}{\partial \phi} - \frac{\partial r_i}{\partial \phi}\left(\frac{\xi}{r_o - r_i(\phi)}\right)\frac{\partial C}{\partial \xi}\right) = \left(\frac{1}{r^2}\right)\left(\frac{\partial}{\partial \xi}\right)\left(\frac{Dr^2}{r_o - r_i(\phi)}\left(\frac{\partial C}{\partial \xi}\right)\right) \tag{35}$$

$$r = r_i(\phi) + \xi\left(r_o - r_i(\phi)\right) \tag{36}$$

$$\frac{\partial \phi}{\partial t} \approx i_{Si-Gr,\, intf} A_{Si}\left(\frac{M_{Si}}{F.m_{Si}}\right) \tag{37}$$

$$i_{Si-Gr,\, intf} = N_{Si-Gr,\, intf} . F \tag{38}$$

The above formulation normalizes the spatial domain, effectively fixing the moving boundary. The influence of the boundary motion is absorbed into the diffusion equation through additional source terms, $\frac{\partial \phi}{\partial t}$ & $\frac{\partial r_i}{\partial \phi}$. The term $\frac{\partial r_i}{\partial \phi}$ is obtained from experimental fits, while $\frac{\partial \phi}{\partial t}$ is approximated as a function of the chemical flux at the silicon–graphite interface similar to experimental formulation for silicon stoichiometry evolution reported in Schmidt H et al. [12].

# 3. Results and discussion

## 3.1 Exploring the Effect of the Chemical Potential Difference

The effect of introducing a chemical reaction between lithiated graphite and silicon is explored in this section.

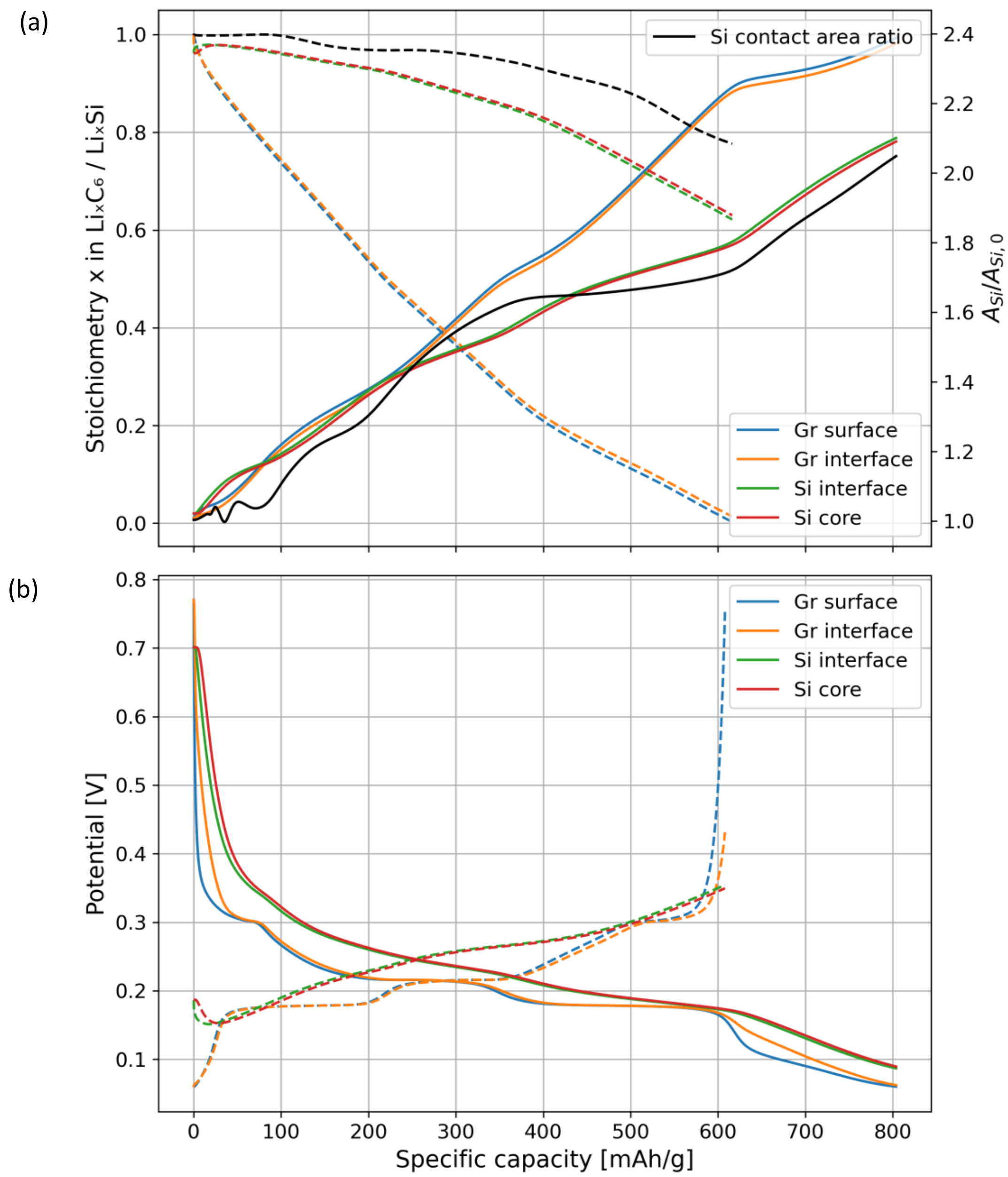

Fig. 5 Concentration (a) & potential (b) distribution in silicon-graphite core-shell structure; lithiation (—), delithiation (– –).

Figures 5(a) and 5(b) illustrate the evolution of lithium concentration and potential distributions in the silicon-graphite core-shell structure during (de)lithiation, highlighting the emergence and impact of chemical potential differences between the two materials.

In Figure 5(a), the normalized lithium concentrations during lithiation (solid lines) show that graphite lithiates more quickly than silicon in the early stages. This results in higher lithium concentration in the graphite shell than in the silicon core, a consequence of the limited chemical reaction rate at the graphite-silicon interface, which restricts lithium transfer into silicon. As lithiation progresses, lithium gradually accumulates in the silicon core, and the interfacial concentration gap narrows slightly.

During delithiation (dashed lines), the behaviour is reversed: silicon retains a higher lithium concentration than graphite throughout much of the process. This again stems from kinetic limitations at the interface. Here, the chemical reaction rate from silicon is lower, causing lithium to exit the graphite shell more rapidly.

These spatial gradients are directly linked to the potential distributions shown in Figure 5(b). During lithiation, the potential in the silicon core remains higher than in graphite due to the lower lithium concentration and the material's intrinsic open-circuit voltage profile. A steep potential drop occurs at the interface, corresponding to the chemical potential difference that drives interfacial transport. As lithium insertion continues, this potential difference decreases.

During delithiation (dashed lines), although silicon retains a higher lithium concentration than graphite for much of the process due to slower chemical reaction kinetics, the local potential in silicon remains higher than that in graphite. This is consistent with the lithium-rich state of silicon and its corresponding open-circuit potential. However, as delithiation nears completion, the graphite shell undergoes rapid deintercalation, especially at the surface. This is reflected in Figure 5(b) by the sharp increase in graphite's surface potential, which begins to dominate and approaches the cut-off voltage. This behaviour signifies the graphite shell reaching its lower stoichiometric limit, at which point it cannot sustain further deintercalation without a substantial voltage rise. As the lithium concentration in the graphite cannot drop below zero the chemical reaction rate cannot be made to go any faster. Therefore, if the electrochemical reaction needs to be faster than this to sustain the current then the cell voltage will collapse almost immediately and the cell will stop discharging, resulting in the lower useable capacities seen experimentally.

Finally, the black line in Figure 5(a) shows the increasing/decreasing contact area ratio $\frac{A_{Si}}{A_{Si,0}}$, which reflects the expansion/contraction of the silicon core. This change modulates the effective interfacial flux by altering the available surface area for the chemical reaction and thus plays a role in both lithiation and delithiation dynamics. For example, during lithiation (solid lines) in Figure 5(a), within the specific capacity range of 0–300 mAh/g, the contact area between silicon and graphite increases rapidly, enabling both materials to increase lithium concentration at comparable rates. The increase of lithium flux into silicon facilitates continued lithium flux into graphite by keeping the concentration at the interface low. In the subsequent range of 300–600 mAh/g, lithium intercalates into graphite faster than the lithium flux into silicon from the chemical reaction, the contact area growth therefore begins to plateau, exacerbating the difference. Beyond 600 mAh/g, the potential difference between graphite and silicon is large enough to drive sufficient lithium flux that the contact area resumes its increase, leading to a partial reconvergence of rates of lithium concentration increase, as the chemical reaction between graphite and silicon becomes more favourable again.

## 3.2 Parameter Sensitivity Analysis

### 3.2.1 Silicon and Graphite Diffusion Coefficients

Table 1. Diffusivity parameters reported in literature for silicon and graphite.

| Parameter | Value & Ref |
|---|---|
| Diffusivity of Li in graphite, $D_{Gr}$ ($m^2/s$) | $3.9\times10^{-14}$[13], $3.3\times10^{-14}$[14], $2\times10^{-14}$[15] $3.3\times10^{-9}$[16] |
| Diffusivity of Li in silicon, $D_{Si}$ ($m^2/s$) | $3\times10^{-16}$[17], $10^{-12}$[15] $1.67\times10^{-14}$[16] $10^{-16}$ |

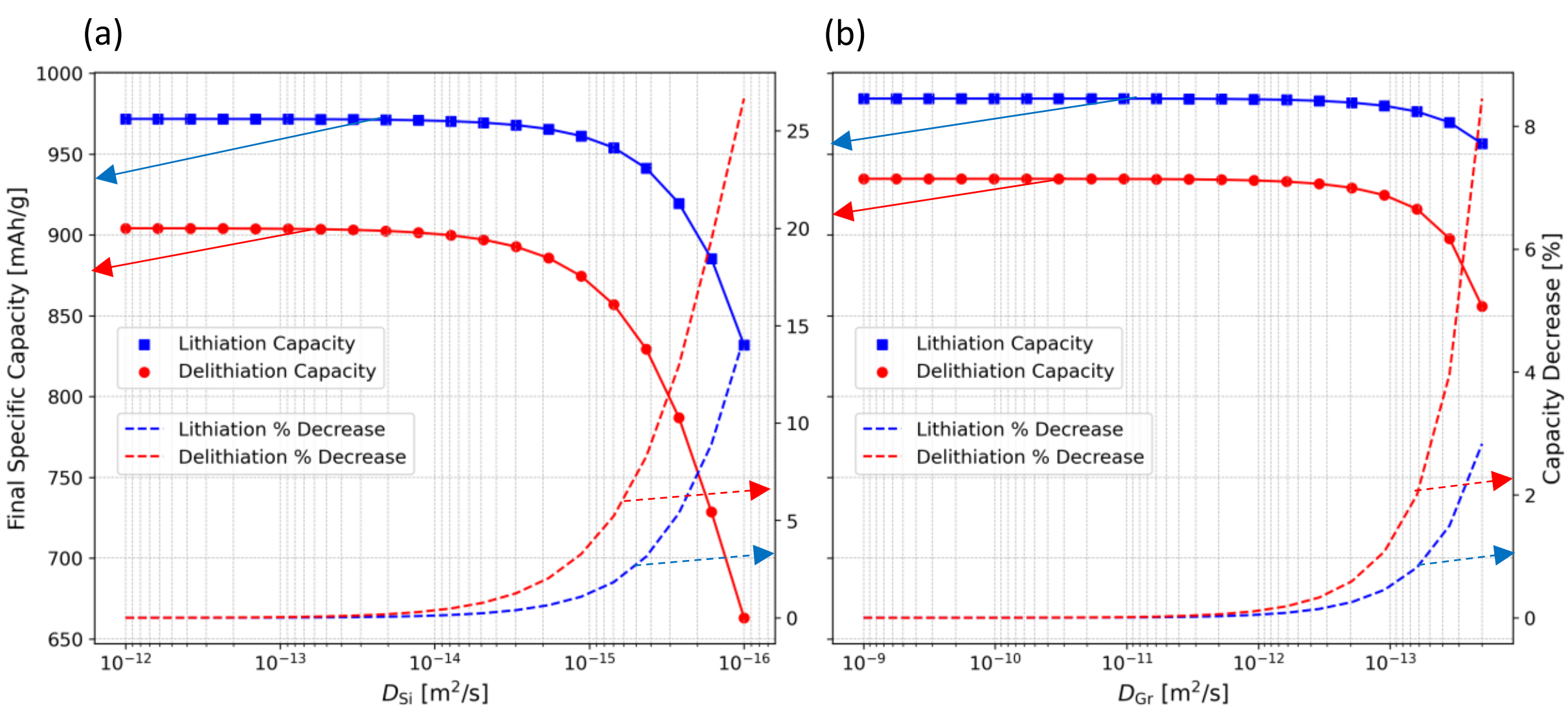

Fig. 6 Effect of varying diffusion coefficients for silicon (a) and graphite (b) on (de)lithiation capacities; red/blue (●/■) = delithiation/lithiation capacity, red/blue (– –) = % decrease in delithiation/lithiation capacity.

From the plots in Figure 6, the impact of lithium diffusivity is far more pronounced in silicon than in graphite. This trend aligns with the diffusivity ranges reported in Table 1. While graphite diffusivity in the literature spans from $2\times10^{-14}$ to $3.3\times10^{-9}$ m²/s, silicon values range from $10^{-16}$ to $10^{-12}$ m²/s. Although there is some overlap, the highest value reported for graphite is approximately 10 million times higher than the lowest value for silicon. On balance, this supports the assumption that lithium diffusion in silicon is significantly slower than in graphite.

As a result, even small changes in silicon's diffusivity within its expected range can substantially affect lithium distribution and the predicted electrochemical response in the model. In contrast, graphite remains relatively insensitive to diffusivity changes over its range. If graphite were artificially assigned diffusivity values characteristic of silicon, similar transport limitations would emerge, reinforcing that the diffusivity parameter must be carefully selected to reflect realistic material behaviour. This highlights the importance of informed parameterization when simulating lithium transport, particularly in composite electrode systems where one phase exhibits orders-of-magnitude lower mobility.

### 3.2.2 Silicon Particle Radius and Graphite Shell Thickness

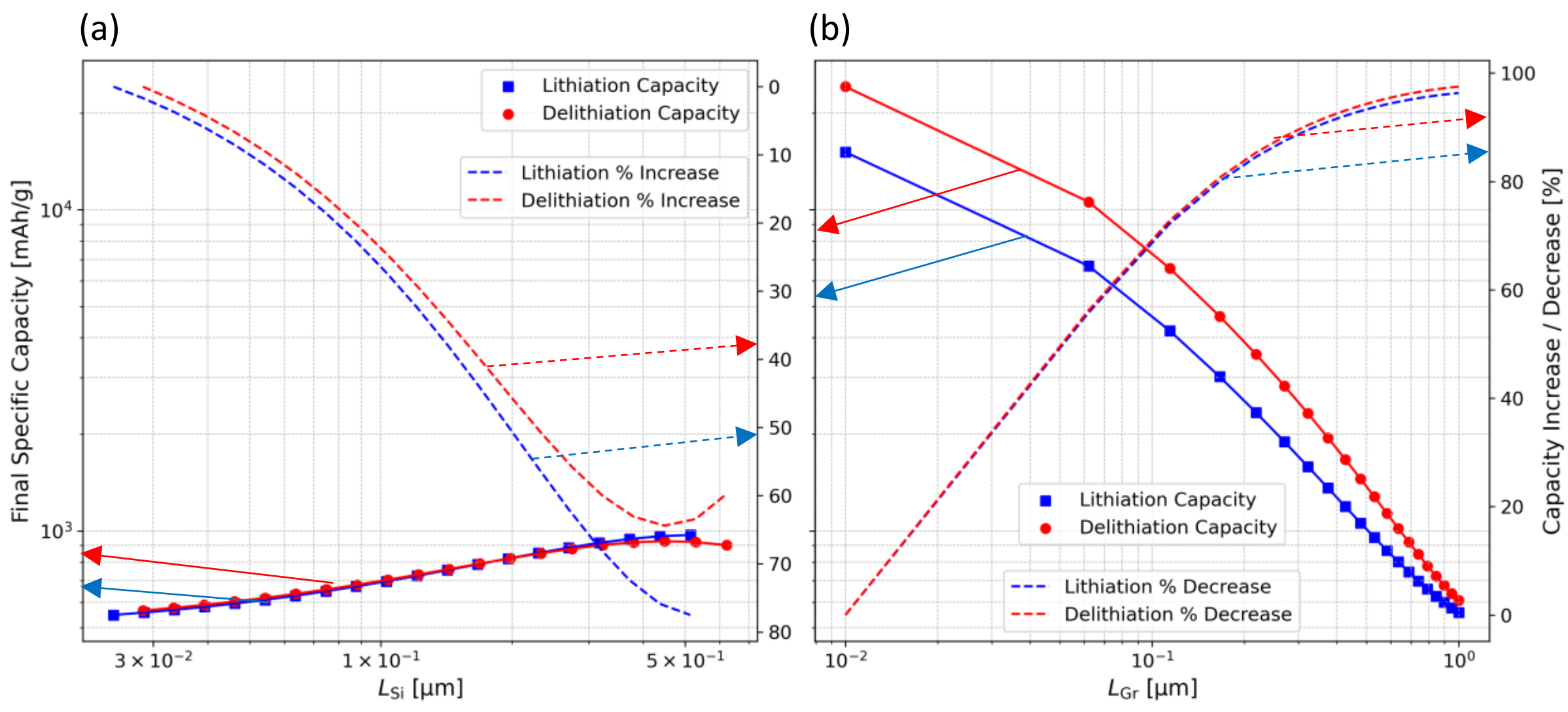

Fig. 7 Effect of varying silicon particle radius (a) and graphite shell thickness (b) on (de)lithiation capacities; red/blue (●/■) = delithiation/lithiation capacity, red/blue (– –) = % increase (a) / decrease (b) in delithiation/lithiation capacity.

Simulations with increasing graphite shell thickness reveal that while the shell initially serves to buffer lithium flux, excessive thickness leads to a noticeable drop in overall capacity as depicted in Figure 7(b). This is primarily due to the diminished lithium flux to the silicon core, resulting in its under-utilisation. A thicker graphite shell increases the diffusion path length and creates a more dominant lithium sink in the shell itself, thereby delaying or limiting lithium transport to the underlying silicon. Consequently, silicon's contribution to the total capacity becomes increasingly suppressed, emphasizing the need to carefully balance graphite shell thickness to optimize both mechanical stability and electrochemical performance.

Figure 7(a) shows varying silicon core thickness. While capacity initially increases with thickness due to greater lithium storage, the benefit reaches a maximum and then begins to decline. This non-monotonic behaviour is attributed to the chemical reaction rate at the graphite–silicon interface becoming a bottleneck. As the silicon core becomes thicker, lithium must traverse a longer path through the silicon, but the chemical reaction rate at the interface governing lithium flux into silicon does not increase accordingly. This mismatch leads to underutilization of the deeper silicon regions and increased lithium accumulation near the interface, reducing overall capacity. Therefore, beyond a critical silicon thickness, the system becomes kinetically limited rather than storage limited. This trend is further emphasized in Figure 8(b), where the diminishing return and eventual decline in capacity with increasing silicon core thickness is clearly observed. Optimizing the silicon core geometry thus requires a careful balance between available capacity and the rate-limiting effects of interface kinetics and diffusion resistance.

### 3.2.3 Silicon-Graphite Interface Chemical Lithiation Rate

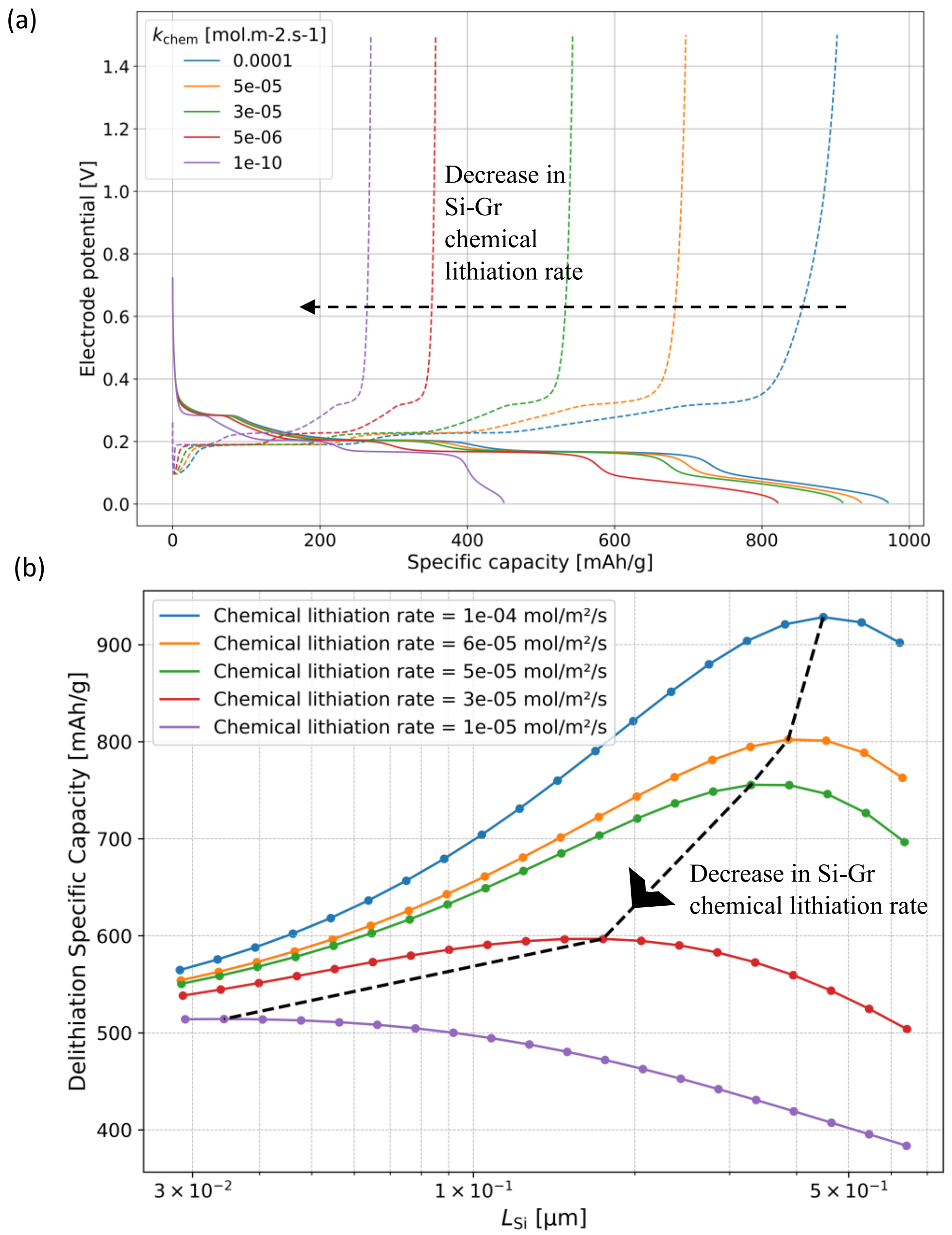

Fig. 8 Effect of varying silicon-graphite interface chemical lithiation rate on capacity (a); lithiation (—), delithiation (– –), and for different silicon particle sizes (b).

Figure 8(a) illustrates that decreasing the chemical lithiation rate across a broad range (from $10^{-3}$ to $10^{-10}$ mol·m$^{-2}$·s$^{-1}$) leads to a marked reduction in overall capacity. This trend underscores the growing influence of interface kinetics as a limiting factor. Further, Figure 8(b) demonstrates how this kinetic limitation manifests more subtly as diminishing returns with increasing silicon core thickness as highlighted in the previous section for Figure 7(a), particularly within the intermediate lithiation rate range of $10^{-4}$ to $10^{-5}$ mol·m$^{-2}$·s$^{-1}$. As the chemical lithiation rate decreases, the optimal silicon core thickness beyond which additional silicon no longer contributes meaningfully to capacity shifts to smaller values. This is emphasized by the black dotted line in Figure 8(b), which tracks the reduction in the optimal (i.e., maximum capacity-yielding) silicon particle size as lithiation rate declines. This behaviour arises because lower interface reaction rates exacerbate the kinetic bottleneck at the graphite–silicon interface, restricting how rapidly lithium can enter the silicon domain. While a thicker silicon core theoretically provides more storage, poor interface kinetics prevent deep lithiation, leading to underutilization and reduced performance. Therefore, these results highlight the critical role of interface kinetics in determining the accessible capacity of the silicon and optimizing the electrode structure accordingly.

Although the chemical lithiation rate is a challenging parameter to measure directly and is not readily available in the literature, its strong influence on capacity behaviour, as demonstrated above, highlights the importance of accurate calibration. Given its sensitivity and critical role in capturing the rate-limiting interface reaction, fitting this parameter to experimental data therefore becomes essential for achieving realistic and predictive modelling of silicon–graphite core-shell composite electrodes.

## 3.3 Importance of CV Hold

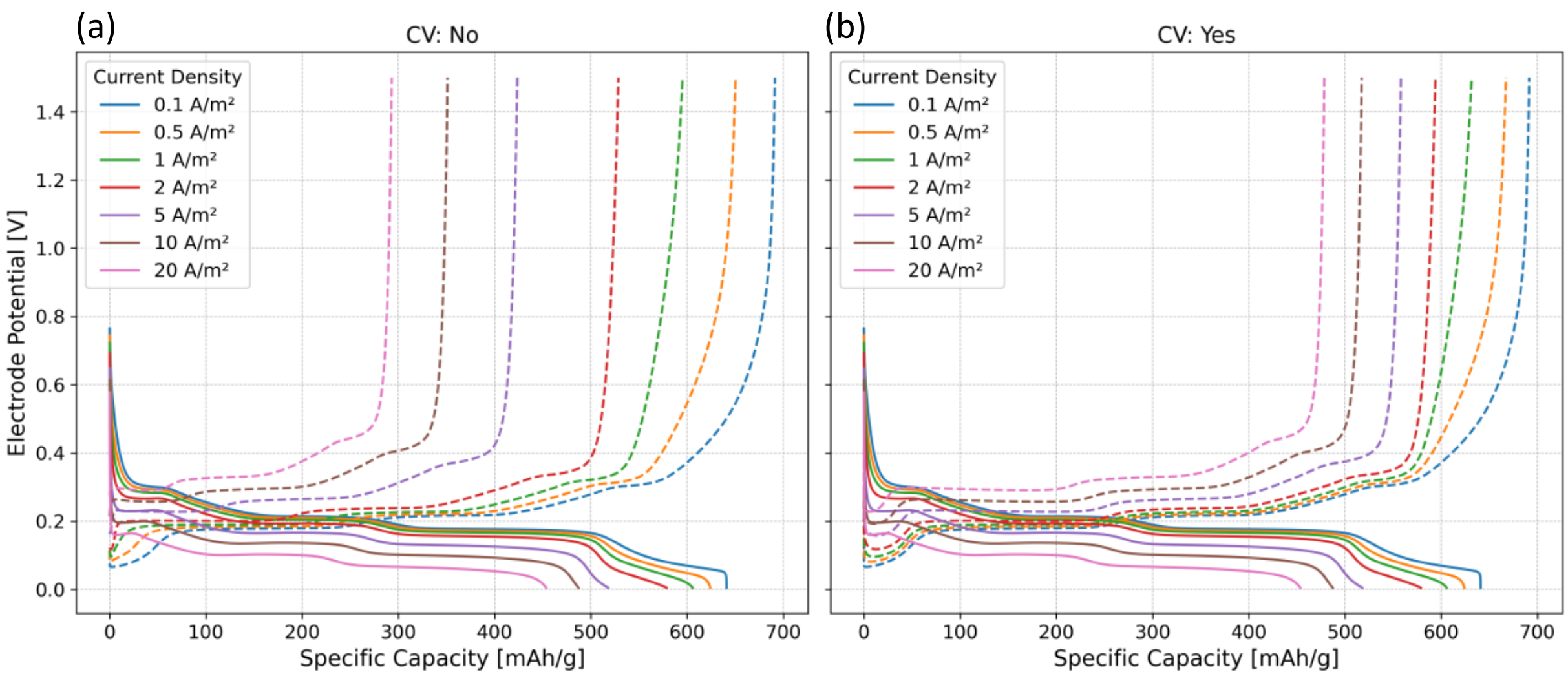

Fig. 9 (De)lithiation without (a) and with (b) constant-voltage hold (CV); lithiation (—), delithiation (– –).

Figure 9 illustrates the impact of including a constant-voltage (CV) hold during lithiation on the subsequent delithiation behaviour. Lithiation (solid lines) is followed by delithiation (dashed lines), with the CV step applied immediately after the lithiation phase. In the absence of a CV hold (left panel), the cell exhibits significant capacity loss during delithiation, especially at higher C-rates (applied current densities). In contrast, the inclusion of a CV step (right panel) substantially mitigates this capacity fade. This improvement is primarily due to the re-balancing of lithium-ion distribution during the CV period, which enables further lithiation across the silicon–graphite interface. This results in greater utilization of the silicon domain, which would otherwise remain partially unlithiated. Without a CV hold, this redistribution cannot occur, leading to prematurely terminated lithiation and lower accessible capacity. These findings underscore the importance of incorporating a CV hold during the charge phase of cells incorporating graphite-coated silicon-based micro- or nanostructures. However, it is equally important to tailor the charge protocol, including the duration of the CV step to the specific operating conditions and performance requirements. In these simulations, for instance, the CV hold duration at the highest simulated current density (10 A/m²) is approximately 1.4 hours, highlighting the need to balance enhanced utilization with practical charging times.

This trend is also evident in Figure 10 below, where the impact of the CV hold becomes more significant with decreasing graphite shell thickness, especially at higher current densities. Thinner graphite shells are deintercalated more rapidly, particularly under high-rate conditions, since the graphite surface concentration directly dictates the cell voltage. As a result, the surface potential quickly approaches the cut-off voltage, prematurely terminating the charge process (left). The CV hold mitigates this effect by allowing additional time for lithium redistribution and interfacial lithiation into the silicon core, thereby enhancing overall capacity (right). This highlights the essential role of the CV hold in ensuring effective utilization of the silicon domain when structural limitations accelerate graphite depletion.

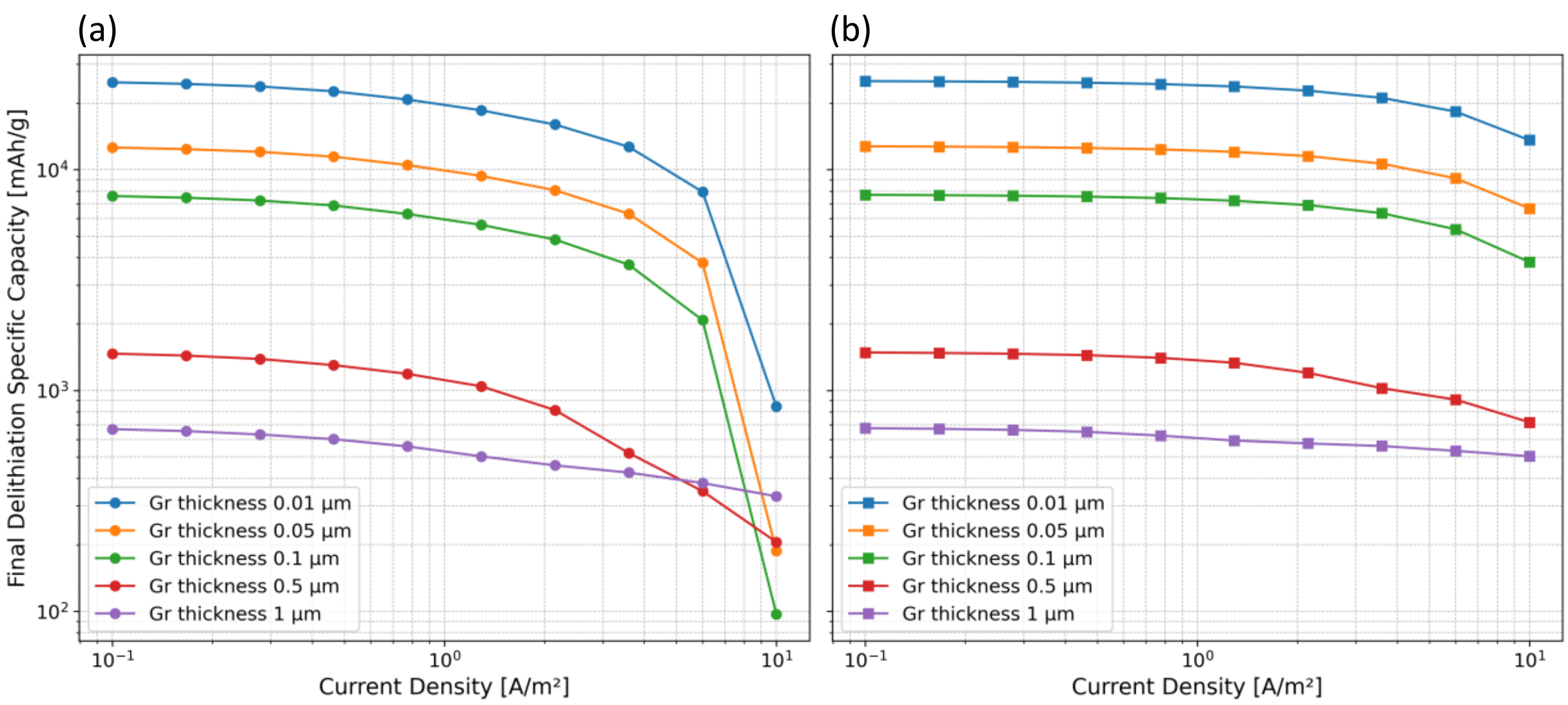

Fig. 10 Effects of varying graphite shell thickness and current density on overall delithiation capacity without CV hold (a) and with CV hold (b).

## 3.4 Model Qualitative Assessment (C-Rate Limitations)

### 3.4.1 Half-Cell C-Rate Limitation Results

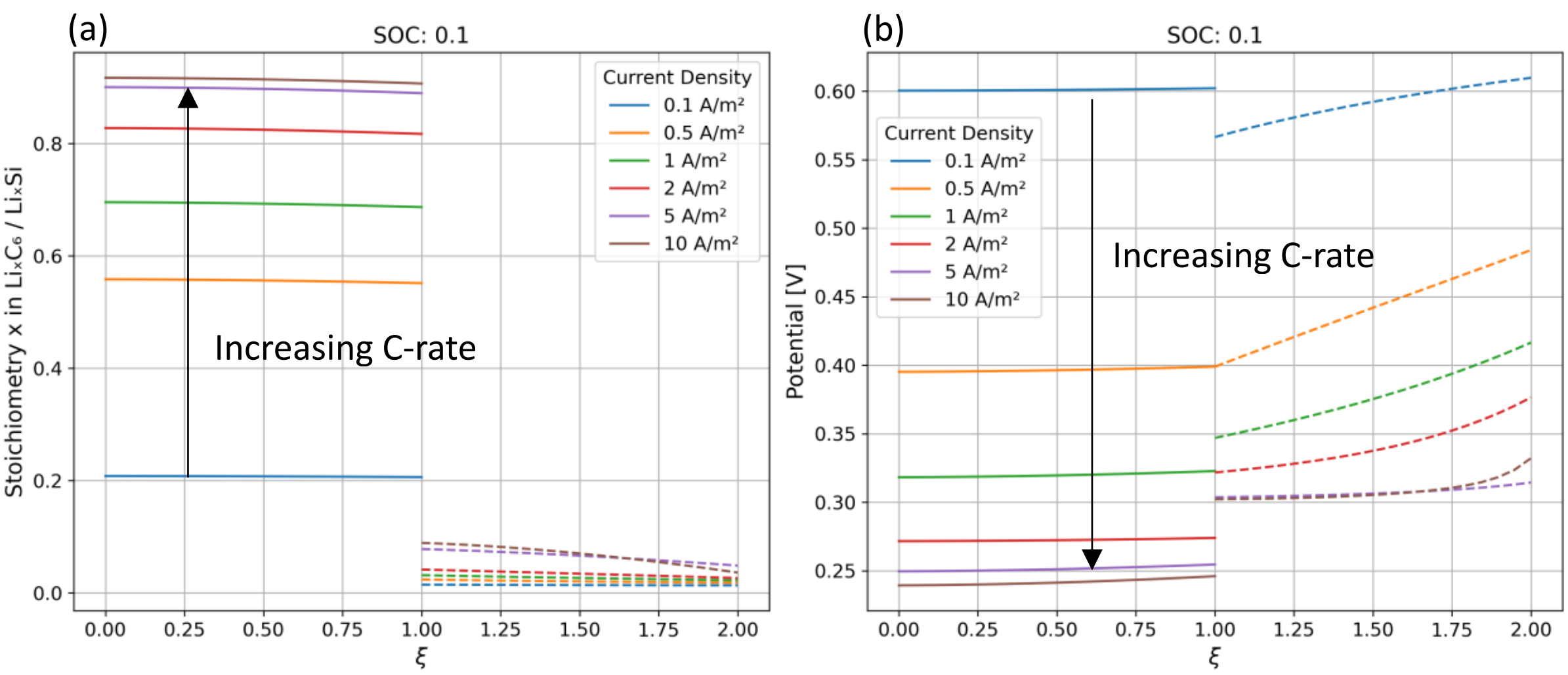

Fig. 11 Spatial variation of $Li^+$ concentration (a) and potential (b) with respect to applied current density during delithiation; silicon core: $0 \le \xi \le 1$ (—), and graphite shell: $1 \le \xi \le 2$ (– –).

Figure 11 illustrates the impact of C-rate (applied current density) on the spatial distribution of normalized $Li^+$ concentration and potential across the electrode domains. The lithium concentration is normalized within each material (silicon and graphite), and the spatial domain is scaled such that the silicon core spans $0 \le \xi \le 1$ and the graphite shell spans $1 \le \xi \le 2$. Focusing on the silicon region (left panel), it is evident that higher C-rates result in lower utilization of silicon during delithiation. This is reflected by the elevated $Li^+$ concentration retained in the silicon at higher current densities, suggesting incomplete lithium extraction. These snapshots correspond to a state of charge (SOC) of 0.1, near the end of the delithiation process, providing a clear comparison of lithium removal efficiency.

The right panel shows the corresponding spatial potential profiles. As the C-rate increases, the potential within the silicon region decreases in line with the $Li^+$ concentration, indicating reduced activity. Meanwhile, the overall cell potential becomes increasingly governed by the graphite surface potential at $\xi = 2$. At high C-rates, this surface potential rises sharply due to rapid deintercalation of $Li^+$, eventually reaching the cut-off voltage. This behavior is consistent with the trends discussed in Section 3.1, where the difference in chemical potential between the silicon core and graphite shell was first introduced.

The effects of C-rate limitation can be further studied with varied silicon core particle sizes.

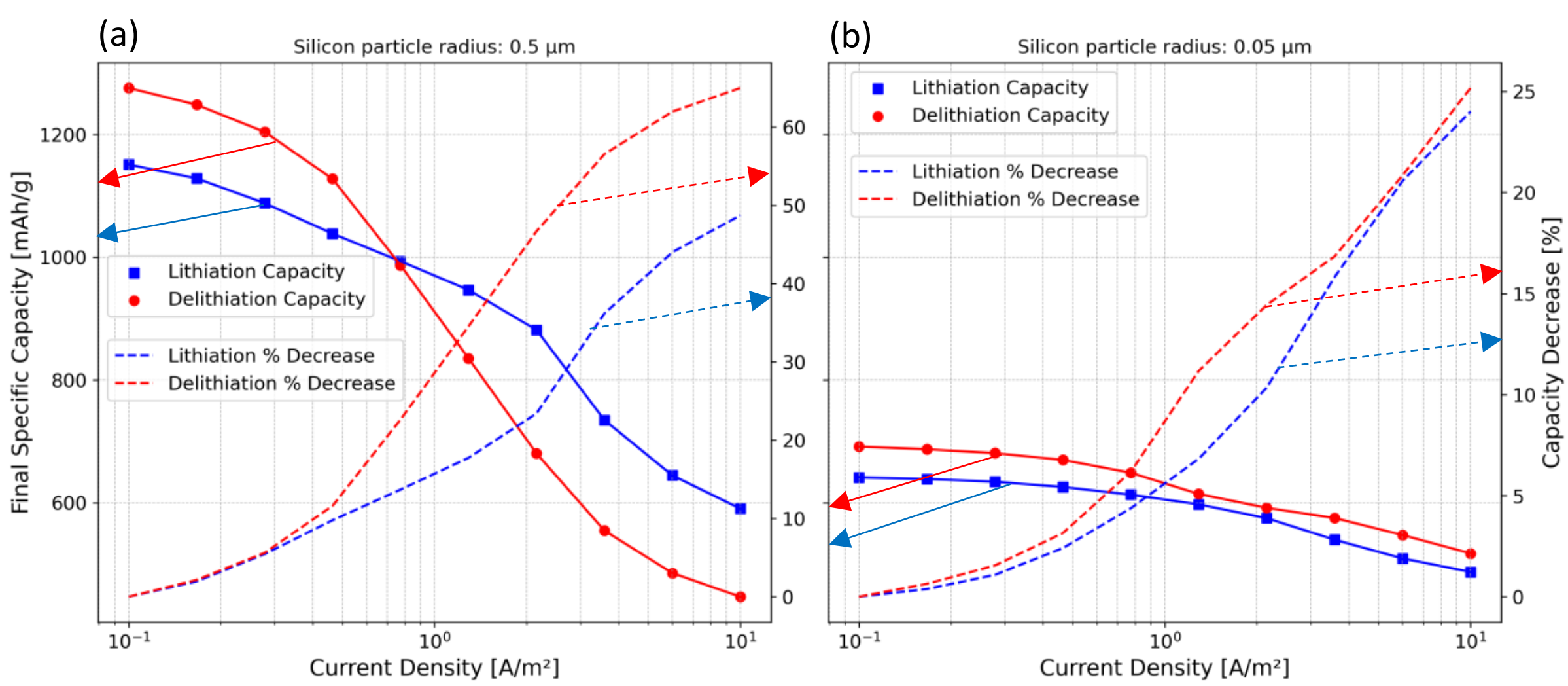

Fig. 12 Capacity fade with C-rate for silicon particles sizes of 0.5 μm (a) and 0.05 μm (b); red/blue (●/■) = delithiation/lithiation capacity, red/blue (– –) = % decrease in delithiation/lithiation capacity.

The results in Figure 12 clearly demonstrate that capacity retention at higher C-rates is significantly better for smaller silicon core particle sizes (0.05 μm, right panel) compared to larger ones (0.5 μm, left panel). At the highest tested C-rate, the larger silicon particles exhibit a capacity reduction of approximately 60%, while the smaller particles show a more modest reduction of about 25%. Despite this, the larger silicon particles deliver a higher specific capacity at low C-rates (~1300 mAh/g) compared to the smaller particles (~700 mAh/g), owing to their greater lithium storage volume. However, this advantage diminishes rapidly as the current density increases due to stronger diffusion limitations and kinetic bottlenecks in the larger particles. The longer diffusion paths in the bulk of the larger particles lead to incomplete lithiation/delithiation and pronounced lithium concentration gradients, reducing overall efficiency under high-rate operation. In contrast, the smaller silicon particles exhibit more favorable kinetics and more uniform lithium distribution, enabling better utilization of the active material even under demanding C-rate conditions. These findings highlight the critical trade-off between achievable capacity and rate performance and emphasize the importance of optimizing silicon particle size for target application requirements in these graphite-coated silicon composite electrodes.

Figure 13 below shows the overall half-cell results with comparison to the experimental results obtained from [7] and [10] for a larger and smaller silicon core size respectively. The results below highlight the ability of the model to qualitatively capture the effects of C-rate limitations and influence of particle sizes. The experimental results are smoother, compared to a sharper model prediction. The real materials clearly have multiple silicon cores in a complicated carbon matrix, therefore with a wide variety of diffusion lengths and time constants giving smoother curves. In contrast the model has only one particle approximating the entire electrode, with a single graphite shell thickness and silicon core thickness, giving a much sharper result. This could be solved by introducing multiple particles with different graphite and silicon thicknesses or changing the geometry and assumptions in the model itself.

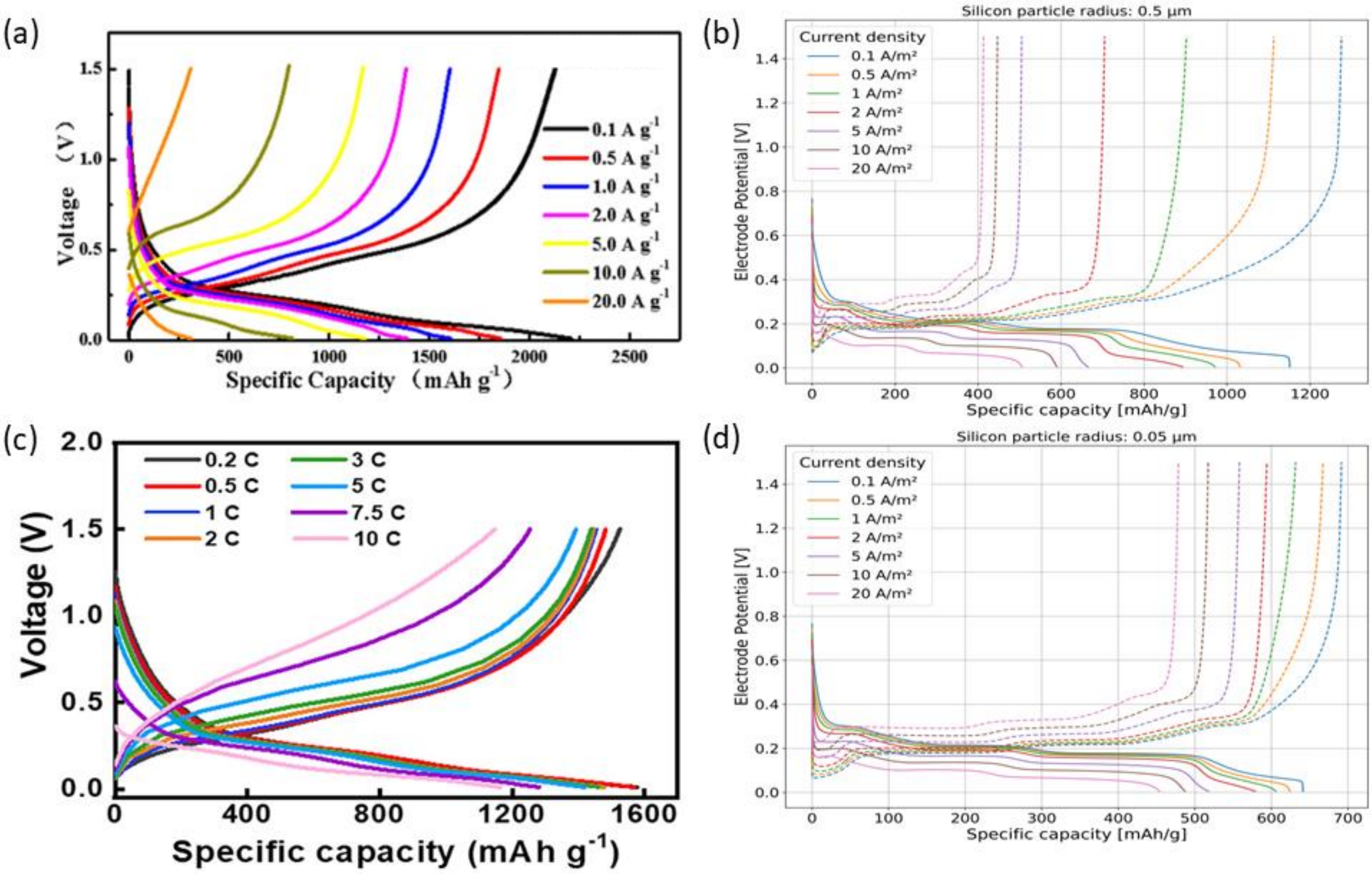

Fig. 13 Qualitative comparison of C-rate limitation between model predictions (b), (d); lithiation (—), delithiation (– –), and experimental results from [7], (a) and [10], (c) for both large and small silicon core particle sizes respectively. Reproduced with permission from the journals as cited in [7, 10].

### 3.4.2 Full-Cell C-Rate Limitation Results

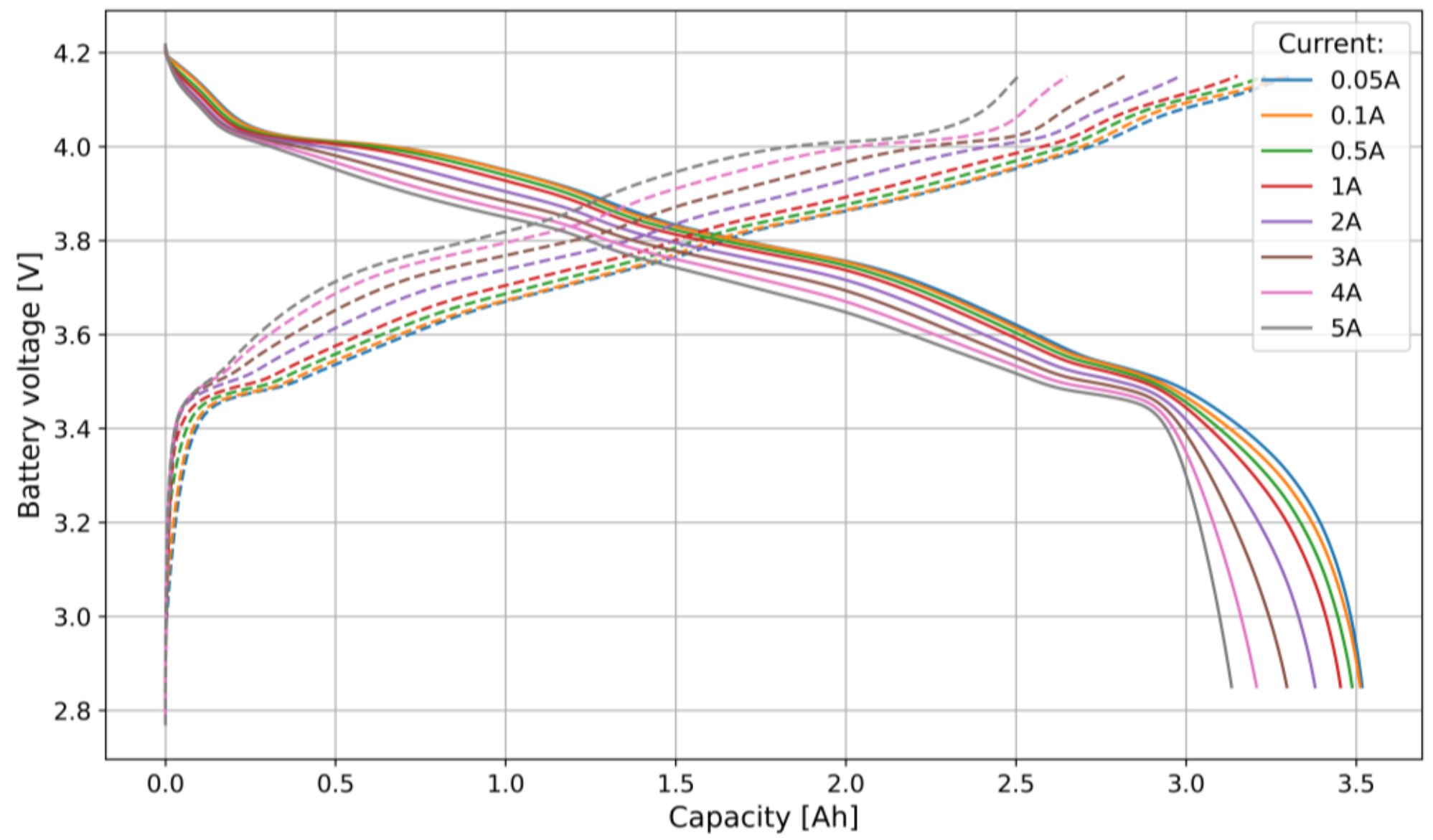

Fig. 14 Full-cell SPM model (Si-C core–shell anode & NMC cathode) results for C-rate limitations; discharge (—), charge (– –).

The model is also extended to a full-cell configuration, as shown in Figure 14, where the Si-C core–shell anode is coupled with an NMC cathode. Both electrodes are modelled using the Single Particle Model (SPM) framework, with parameter sets derived from Chen et al. [18] for the commercial LGM50 cell. The full-cell simulation successfully captures the characteristic C-rate limitations imposed by the Si–C anode, highlighting the strong kinetic and transport constraints associated with its core–shell architecture. These limitations manifest not only as reduced accessible capacity at higher C-rates, but also as a pronounced decrease in round-trip energy efficiency, driven primarily by increased overpotentials in the Si–C anode. At moderate to high C-rates, the resulting voltage hysteresis and polarization losses lead to substantial irreversible energy dissipation, implying elevated heat generation that could pose additional challenges for thermal management in practical cell operation. Future improvements could involve upgrading to more detailed full-cell models, such as the Single Particle Model with electrolyte (SPMe) or the Doyle-Fuller-Newman (DFN) model, to better resolve electrolyte effects and interfacial dynamics. Additionally, incorporating degradation mechanisms, including silicon particle cracking, lithium plating, and SEI growth would enable a more comprehensive study of long-term cycling behaviour at the full-cell scale, even though the following section already addresses cracking in the Si-C half-cell context.

## 4. Introducing Cracking (A Degradation Study)

To study the long-term mechanical degradation of the Si-C anode, the electrochemical model presented in Section 2 is extended by incorporating a simple cracking model, following the approach proposed by O'Kane et al. [19]. This extension introduces mechanical stress analysis, crack propagation dynamics, and the resulting loss of active material (LAM). The governing equations for particle stress (39-41), crack growth (42), and LAM (43) evolution are provided below:

$$\sigma_r = \frac{2\Omega E}{1-\nu}\left(\bar{c}(R_i) - \bar{c}(r)\right), \quad \text{simplification: } \sigma_r = 0 \tag{39}$$

$$\sigma_t = \frac{\Omega E}{1-\nu}\left(2\bar{c}(R_i) + \bar{c}(r) - \frac{\tilde{c}}{3}\right) \tag{40}$$

$$\bar{c}(r) = \frac{1}{3r^3}\int_0^r \tilde{c}\, r^2\, dr, \quad \tilde{c} = c - c_{ref}, \quad c_{ref} = 0 \tag{41}$$

$$\frac{dl_{\mathrm{cr}}}{dN} = k_{\mathrm{cr}} \cdot \frac{1}{t_0} \cdot \sigma_t^{b_{\mathrm{cr}}}\left(\sqrt{\pi l_{\mathrm{cr}}}\right)^{m_{\mathrm{cr}}}, \quad \text{for } \sigma_t > 0 \tag{42}$$

$$\frac{d\varepsilon_a}{dt} = \frac{b}{t_0}\left(\frac{\sigma_{h,\max} - \sigma_{h,\min}}{\sigma_c}\right)^{m_2}, \quad \text{for } \sigma_{h,\min} > 0 \tag{43}$$

This simplified cracking model is based on the following key assumptions:

- Radial particle stresses ($\sigma_r$) are considered negligible in contributing to crack formation, with hoop (tangential) ($\sigma_t$) stress being the dominant factor.
- There is no mechanical stress–strain coupling between the graphite shell and silicon core; each material is treated independently in terms of stress evolution and crack propagation.
- Crack propagation and the resulting loss of active material (LAM) are evaluated locally at distinct interfaces, with each interface possessing its own set of cracks and corresponding localized LAM ($LAM_{local}$).
- The model accounts for three primary interfaces: (i) the external surface of the graphite shell, (ii) the graphite-side of the silicon–graphite interface, and (iii) the silicon-side of the silicon–graphite interface.

These assumptions simplify the mechanical coupling while still capturing the localized nature of stress-induced degradation in the composite particle structure.

For a more physically accurate representation of stress–strain coupling in core–shell particle architectures, the model proposed by Liu et al. [20] can be employed. Their model was originally developed for a hollow $LiCoO_2$ particle undergoing a phase change, which shares structural similarities with the silicon–graphite core–shell configuration used in our study, namely, a distinct interface between two regions with different mechanical and electrochemical properties. While their framework captures the mechanical interactions across this interface with greater fidelity, it introduces significant complexity. As such, it is not implemented in the present work. Nevertheless, future studies could benefit from adopting such a model to investigate the coupled stress evolution and its impact on degradation mechanisms more rigorously. The parameters used for the simple cracking model are shown in Table S3 and variables solved for in Table S4.

The results obtained from the simple cracking model are presented in Figure 15 below, alongside experimental data from [8] and [9].

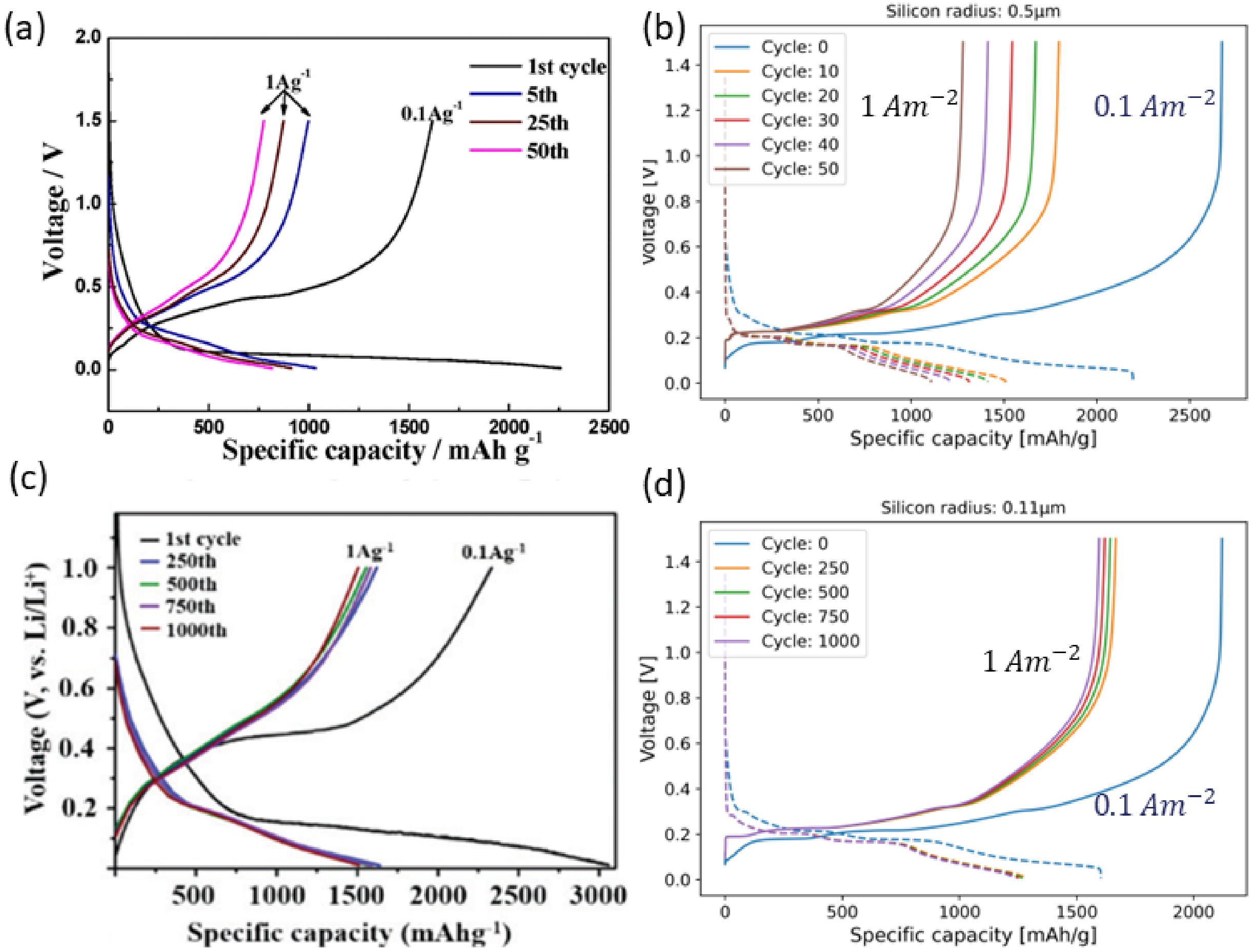

Fig. 15 Cycling-degradation results comparison experiment from [8], (a) and [9], (c) vs model (b) and (d); lithiation (– –), delithiation (—). Reproduced with permission from the journals as cited in [8, 9].

The results obtained from the simple cracking model qualitatively reproduce the cycling-induced degradation trends observed in experimental studies. As shown in Figure 15, the model captures the characteristic capacity fade over repeated cycles, which aligns well with the experimental data from [8] and [9]. In particular, the extent of degradation is shown to be influenced by the silicon particle size, with larger particles exhibiting more severe capacity loss due to their greater mechanical stress and susceptibility to cracking. This supports the hypothesis that mechanical degradation, primarily driven by volume expansion and stress-induced cracking in silicon is a dominant mechanism of capacity fade in such electrode architectures.

## 5. Parameters and Variables Summary

Table S1: Model Parameters

| **Parameter** | **Symbol** | **Unit** | **Value** |
|---|---|---|---|
| Faraday constant | $F$ | C mol$^{-1}$ | 96485 |
| Universal gas constant | $R$ | J mol$^{-1}$ K$^{-1}$ | 8.314 |
| Temperature | $T$ | K | 298.15 |
| Maximum Li concentration (Gr) | $C_{Gr}^{\max}$ | mol m$^{-3}$ | 30555 |
| Maximum Li concentration (Si) | $C_{Si}^{\max}$ | mol m$^{-3}$ | 311000 |
| Charge transfer coefficient | $\alpha$ | - | 0.5 |
| Intercalation rate constant (Gr) | $k^{Gr}$ | m$^{2.5}$ mol$^{-0.5}$ s$^{-1}$ | $5.031\times10^{-11}$ |
| Diffusion coefficient (Gr) | $D_{Gr}$ | m$^{2}$ s$^{-1}$ | $3.9\times10^{-14}$ |
| Diffusion coefficient (Si) | $D_{Si}$ | m$^{2}$ s$^{-1}$ | $1.67\times10^{-14}$ |
| Backward reaction rate (Si→Gr) | $k_b$ | mol m$^{-2}$ s$^{-1}$ | $1.0\times10^{-4}$ |
| Forward reaction rate (Gr→Si) | $k_f$ | mol m$^{-2}$ s$^{-1}$ | **Calculated from** (3, 20-23) |
| Initial silicon particle radius | $R_0$ | m | $0.5\times10^{-6}$ |
| Initial contact area (Si-Gr) | $A_{Cont,0}$ | m$^{2}$ | **Evaluated from $\frac{V}{V_0}$ data** |
| Applied current density | $i$ | A m$^{-2}$ | 1.0 |
| Silicon volume ratio | $\frac{V}{V_0}$ | - | **Obtained from experimental results [11]** |

| Table S2: Variables to Solve For | | | |
|---|---|---|---|
| **Variable** | **Symbol** | **Unit** | **Description** |
| Li stoichiometry (Gr) | $X_{Li[Gr]}$ | - | $C_{Gr}/C_{Gr}^{\max}$ (local lithiation in graphite) |
| Li stoichiometry (Si) | $X_{Li[Si]}$ | - | $C_{Si}/C_{Si}^{\max}$ (local lithiation in silicon) |
| Equilibrium potential (Gr) | $E_{eq}^{Gr}$ | V | **Function of** $X_{Li[Gr]}$ (2) |
| Equilibrium potential (Si, lithiation) | $E_{eq,lith}^{Si}$ | V | **Function of** $X_{Li[Si]}$ (5) |
| Equilibrium potential (Si, delithiation) | $E_{eq,delith}^{Si}$ | V | **Function of** $X_{Li[Si]}$ (5) |
| Overpotential (Gr) | $\eta^{Gr}$ | V | Overpotential at graphite-electrolyte interface |
| Exchange current density (Gr) | $i_0^{Gr}$ | A $m^{-2}$ | **Function of** $X_{Li[Gr]}$ (8-9) |
| Surface potential (Gr) | $\phi^{Gr}$ | V | Electrode potential (10) |
| Cell voltage | $V$ | V | Output voltage (11) |
| Li flux (Gr/Si interface) | $N_{Gr/Si}$ | mol $m^{-2}$ $s^{-1}$ | Net Li transfer rate (17-19) |
| Contact area (Si-Gr) | $A_{Cont}$ | $m^2$ | Dynamic area due to Si expansion (26) |
| Silicon core thickness | $d_{Si}$ | m | Effective diffusion length (28) |

Table S3: Cracking Parameters

| **Parameter** | **Symbol** | **Unit** | **Material** | **Value** |
|---|---|---|---|---|
| Critical hoop stress (Si) | $\sigma_c^{\mathrm{Si}}$ | Pa | Silicon | $3.0 \times 10^{9}$ |
| Crack growth coefficient (Si) | $k_{\mathrm{cr}}^{Si}$ | - | Silicon | $2.5 \times 10^{-20}$ |
| Crack growth exponents (Si) | $b_{cr}^{Si}, m_{cr}^{Si}$ | - | Silicon | 1.125, 2.18 |
| LAM coefficient & exponent (Si) | $b_{Si}, m_2^{Si}$ | - | Silicon | $2.7778 \times 10^{-7}$, 2.0 |
| Young's modulus (Si) | $E_{Si}$ | Pa | Silicon | $8.0 \times 10^{10}$ |
| Poisson's ratio (Si) | $\nu_{Si}$ | - | Silicon | 0.28 |
| Partial molar volume (Si) | $\Omega_{Si}$ | $m^3\ mol^{-1}$ | Silicon | $8.18 \times 10^{-6}$ |
| Critical hoop stress (Gr) | $\sigma_c^{Gr}$ | Pa | Graphite | $6.0 \times 10^{7}$ |
| Crack growth coefficient (Gr) | $k_{\mathrm{cr}}^{Gr}$ | - | Graphite | $3.9 \times 10^{-20}$ |
| Crack growth exponents (Gr) | $b_{cr}^{Gr}, m_{cr}^{Gr}$ | - | Graphite | 1.12, 2.2 |
| LAM coefficient & exponent (Gr) | $b_{Gr}, m_2^{Gr}$ | - | Graphite | $2.7778 \times 10^{-7}$, 2.0 |
| Young's modulus (Gr) | $E_{Gr}$ | Pa | Graphite | $1.5 \times 10^{10}$ |
| Poisson's ratio (Gr) | $\nu_{Gr}$ | - | Graphite | 0.3 |
| Partial molar volume (Gr) | $\Omega_{Gr}$ | $m^3\ mol^{-1}$ | Graphite | $3.1 \times 10^{-6}$ |

| Table S4: Cracking Variables | | | | |
|---|---|---|---|---|
| **Variable** | **Symbol** | **Unit** | **Material** | **Description** |
| Hoop stress (Si) | $\sigma_t^{\mathrm{Si}}$ | Pa | Silicon | Tangential stress in Si particles |
| Crack length (Si) | $l_{\mathrm{cr}}^{Si}$ | m | Silicon | Length of cracks in Si |
| Loss of active material (Si) | $LAM_{Si}$ | - | Silicon | Fraction of inactive Si due to cracking |
| Hoop stress (Gr) | $\sigma_t^{Gr}$ | Pa | Graphite | Tangential stress in Gr shell |
| Crack length (Gr) | $l_{\mathrm{cr}}^{Gr}$ | m | Graphite | Length of cracks in Gr |
| Loss of active material (Gr) | $LAM_{Gr}$ | - | Graphite | Fraction of inactive Gr due to cracking |
| Radial stress (both) | $\sigma_r$ | Pa | Both | Negligible (assumed 0 in model) |
| Reference Li concentration | $c_{ref}$ | mol $m^{-3}$ | Both | Stress-free reference state (set to 0) |

## 6. Conclusion

This study presents a physics-based modeling framework to investigate the electrochemical and mechanical behavior of graphite-coated silicon (Si-C) core-shell particles used as anodes (negative electrodes) in lithium-ion batteries. The model introduces a chemical reaction at the graphite-silicon interface which becomes rate limiting under realistic conditions. The overall mechanism can be considered a 4-step electrochemical-diffusion-chemical-diffusion process (EC-D-C-D), compared to a normal model for a single particle being simply a 2-step electrochemical-diffusion (EC-D) process. A detailed analysis was carried out to understand the interplay between particle geometry, charge protocols, and interfacial chemical reaction kinetics, and how these factors collectively influence capacity utilization and degradation. The results demonstrate the strong sensitivity of overall capacity to the thickness of the silicon core and graphite shell, revealing key trade-offs between available lithium storage and interface-limited lithiation kinetics. Specifically, increasing the silicon core thickness initially enhances capacity but eventually leads to a decline due to a bottleneck at the graphite–silicon interface, a result that is further emphasized in parameter sweeps of the interfacial chemical reaction rate.

Additionally, the importance of implementing a constant-voltage (CV) hold step in the charging protocol was highlighted. Simulations show that the CV hold plays a critical role in rebalancing lithium ions across the silicon–graphite interface, especially after higher current densities and with thinner graphite shells, thereby mitigating kinetic limitations and improving capacity utilization. C-rate studies further demonstrate that smaller silicon particles exhibit significantly improved capacity retention under fast charging conditions, despite having lower absolute capacity at low c-rates, suggesting a critical size threshold for minimizing kinetic and mechanical limitations.

A full-cell model coupling the Si-C core-shell anode with an NMC cathode, both modeled using the single-particle model (SPM) and parameterized using data from Chen et al. for the LG M50 cell further validates the core-shell behavior under practical full-cell operation. While this simplified full-cell configuration captures the rate limitations imposed by the anode, future work could benefit from employing more detailed SPMe or DFN models and incorporating degradation processes.

To explore degradation, the electrochemical model was coupled with a simple stress-based cracking model. The degradation model considers crack propagation and loss of active material (LAM) at three key interfaces in the particle: the graphite surface, the graphite–silicon boundary (on both sides), and captures the localized mechanical failure mechanisms arising from stress accumulation due to volume expansion. While simplified, the model predicts qualitative agreement with experimental cycling degradation trends, particularly the increased degradation severity observed with larger silicon particles. This reinforces the hypothesis that cracking, driven by silicon's significant volume expansion, is the primary degradation mechanism in these composite particles.

## 6. Acknowledgement

The research leading to these results has received funding from the Innovate UK HISTORY project (Grant number 10040711). The authors gratefully acknowledge the support and contributions of the PyBaMM development team, whose open-source platform was instrumental in the implementation and simulation of the models presented in this work.

We would also like to thank Harald Schmidt, Bujar Jerliu, Erwin Hüger (Technische Universität Clausthal, Institut für Metallurgie and Clausthaler Zentrum für Materialtechnik), and Jochen Stahn (Laboratory for Neutron Scattering and Imaging, Paul Scherrer Institut) for providing experimental data on silicon (de)lithiation-induced volume expansion, which was directly incorporated into the model to represent the mechanical behavior of silicon more accurately.

## 7. Credit Author Statement

Dharshannan Sugunan: Methodology, Software, Writing – Original Draft, Visualisation. Yang Jiang: Methodology, Software, Writing – Review & Editing. Jia Guo: Methodology, Writing – Original Draft, Writing – Review & Editing. Gregory Offer: Conceptualisation, Methodology, Writing – Review & Editing, Supervision, Project administration, Funding acquisition. Huizhi Wang: Conceptualisation, Supervision, Project administration. Monica Marinescu: Supervision, Project administration.

## 8. Nomenclature and Abbreviations

**Latin Symbols**

|  |  |
|---|---|
| $A$ | Contact area ($m^2$) |
| $A_{Cont}$ | Dynamic Si-Gr interface area ($m^2$) |
| $C$ | Lithium concentration (mol $m^{-3}$) |
| $D_{Gr}, D_{Si}$ | Diffusion coefficients for graphite/silicon ($m^2$ $s^{-1}$) |
| $E_{eq}$ | Equilibrium potential (V) |
| $F$ | Faraday constant (C $mol^{-1}$) |
| $i$ | Current density (A $m^{-2}$) |
| $i_0$ | Exchange current density (A $m^{-2}$) |
| $k_f, k_b$ | Forward/backward reaction rates (mol $m^{-2}$ $s^{-1}$) |
| $N_{Gr/Si}$ | Li flux at interface (mol $m^{-2}$ $s^{-1}$) |
| $R$ | Universal gas constant (8.314 J $mol^{-1}$ $K^{-1}$) |
| $T$ | Temperature (K) |
| $V$ | Cell voltage (V) |
| $X_{Li}$ | Lithium stoichiometry (-) |
| $r_i, r_o$ | Inner/outer radii of silicon core (m) |
| $\xi$ | Normalized spatial coordinate (-) |

**Greek Symbols**

|  |  |
|---|---|
| $\alpha$ | Charge transfer coefficient (-) |
| $\eta$ | Overpotential (V) |
| $\nu$ | Poisson's ratio (-) |
| $\sigma$ | Mechanical stress (Pa) |
| $\Omega$ | Partial molar volume ($m^3$ $mol^{-1}$) |
| $\phi$ | Surface stoichiometry of silicon particle (-) & graphite surface concentration (V) |

**Subscripts/Superscripts**

| | |
|---|---|
| cr | Crack-related |
| eq | Equilibrium |
| Gr | Graphite |
| Si | Silicon |
| lith | Lithiation |
| delith | Delithiation |

**Abbreviations**

| | |
|---|---|
| CV | Constant Voltage |
| DFN | Doyle-Fuller-Newman model |
| Gr | Graphite |
| LAM | Loss of Active Material |
| NMC | Nickel Manganese Cobalt oxide |
| SEI | Solid Electrolyte Interphase |
| Si | Silicon |
| SOC | State of Charge |
| SPM | Single Particle Model |
| SPMe | Single Particle Model with electrolyte |